\newcommand{\titlename}{Group Theory and Representation Theory for Identical Particles}

\documentclass[twoside]{article}

\usepackage{amsmath}
\usepackage{mathtools}
\usepackage{mathabx} 
\usepackage{lineno}
\usepackage[colorlinks]{hyperref}
\usepackage[dvipsnames]{xcolor}

\hypersetup{
linkcolor=BrickRed
,citecolor=Green
,filecolor=Mulberry
,urlcolor=NavyBlue
,menucolor=BrickRed
,runcolor=Mulberry
,linkbordercolor=BrickRed
,citebordercolor=Green
,filebordercolor=Mulberry
,urlbordercolor=NavyBlue
,menubordercolor=BrickRed
,runbordercolor=Mulberry,
pdftitle = {\titlename},
    pdfauthor = {James Daniel Whitfield}
}

\hypersetup{
linkcolor=NavyBlue
,citecolor=NavyBlue
,filecolor=Red
,urlcolor=NavyBlue
,menucolor=Red
,runcolor=Red
,linkbordercolor=NavyBlue
,citebordercolor=NavyBlue
,filebordercolor=Red
,urlbordercolor=NavyBlue
,menubordercolor=Red
,runbordercolor=Red,
pdftitle = {\titlename},
    pdfauthor = {James Daniel Whitfield}
}

\usepackage{fancyhdr}
\usepackage[sectionbib,numbers,sort&compress]{natbib}  
\usepackage{subfig}
\usepackage{bbold}
\usepackage{minitoc}
\usepackage{appendix}

\usepackage{imakeidx}
\makeindex

\usepackage{graphicx}

\usepackage[
  margin=1in,
  includefoot,
  footskip=30pt,heightrounded
]{geometry}

\usepackage{blochsphere}

\usetikzlibrary{arrows.meta}
\graphicspath{{Images/}}
\usepackage{tikz}

\usetikzlibrary{calc}
\usetikzlibrary{quantikz2}
\usetikzlibrary{circuits.logic.US}

\usepackage[RPvoltages]{circuitikz}

\DeclareMathOperator{\Tr}{Tr}

\definecolor{darkgreen}{rgb}{0,0.6,0}

\newcommand{\id}{\mathbf{1}}
\newcommand{\PP}{\mathbf{P}}

\newcommand{\vX}{\mathcal{V}}

\newcommand{\ej}[1]{\ket{\mathbf{\hat e}_{#1}}}
\newcommand{\edj}[1]{\bra{\mathbf{\hat e}_{#1}}}

\newcommand{\ad}{a^\dagger}
\newcommand{\sgn}{\text{sgn}}
\newcommand{\vac}{\ket{vac}}

\newcommand{\reals}{\mathbb{R}}
\newcommand{\cplxs}{\mathbb{C}}

\newcommand{\figref}[1]{Fig.~\ref{#1}}
\renewcommand{\eqref}[1]{Eq.~\ref{#1}}

\renewcommand{\emph}[1]{\textit{#1}\index{#1}}

\newcommand{\grayfootnote}[1]{%
    \renewcommand\thefootnote{\colorbox{black!10}{\textcolor{black}{\arabic{footnote}}}}%
    \footnotemark
    \footnotetext{%
        \colorbox{black!10}{%
            \begin{minipage}{\textwidth-21.30023pt}%
            \textcolor{black}{#1}%
            \end{minipage}%
        }%
    }%
}
\newcommand{\exercise}[1]{\grayfootnote{#1}}

\def\BibTeX{{\rm B\kern-.05em{\sc i\kern-.025em b}\kern-.08em
    T\kern-.1667em\lower.7ex\hbox{E}\kern-.125emX}}
    
\def\@makechapterhead#1{%
  \vspace*{50\p@}
  \Huge \bfseries #1\par\nobreak                 
    \vskip 40\p@
  }
\makeatother

\usepackage[vcentermath]{youngtab}

\title{\titlename}  

\author{James Daniel Whitfield\\ {\small Department of Physics and Astronomy, Dartmouth College} \\ \small
        {Amazon Visiting Academic, AWS Center for Quantum Computing\thanks{JDW holds concurrent appointments at Dartmouth College and at AWS as an Amazon Visiting Academic. This work was performed at Dartmouth College and is not associated with Amazon.}}
        }
\begin{document}

\maketitle

\begin{abstract}
Few, if any, applications of quantum technology are as widely known as the quantum simulation of quantum matter. Consequently, many interesting questions have been sparked at the intersection of condensed matter, quantum chemistry, and quantum computing. Given the common mathematical foundation of these subjects, we walk through the necessary group theory and representation theory serving as background 
in all of these fields. 
Our discussion will include a full development of the mathematics of identical particles and the mechanics of describing systems of identical particles in both first and second quantization schemes. 
This chapter is an offshoot of a larger work that provides a graduate-level introduction to quantum information science. This chapter is being released separately because it is not explicitly focused on quantum information. It has grown beyond a short digression into a full-fledged journey into the symmetries and representations of identical particles that we invite you, the reader, to join. 
\end{abstract}

\tableofcontents

\section{Introduction} 

This chapter contains an introduction to group theory and representation theory. Our focus is on the permutation group because it provides much of the mathematical tools and terminology needed to describe bosons and fermions.
The goal here is a text that is more accessible than rigorous mathematical yet still sufficient to provide the foundational concept underlying the theory of identical particles. 

\section{Identical particles}
Qubits are individuals. They come with labels or with names.  They will have unique features based on location, fabrications, or their memory address. Fermions such as electrons, by contrast, are all the same. Electrons, which have finite mass and angular momentum but no known internal structure, are all created equal. There is no fundamental difference between two fermions of the same species. When they form a multi-particle cloud of probability, they cannot be labeled.  Similarly, bosons are also identical particles. Examples of bosons are photons, phonons (quantized vibrations), and some nuclei.  

Fermions are identical particles that carry the antisymmetric one-dimensional representation of the symmetric group of permutations. Since this representation is one-dimensional, the action of a permutation does not change the quantum probability density matrix. The other one-dimensional representation of the symmetric group irrespective of the number of elements being permuted is the completely symmetric representation which corresponds to bosons.

Understanding and modeling interacting quantum particles is one problem researchers in the fields of chemistry, physics, engineering, and material science face. For example, in molecular physics, one of the most important tasks is to quickly calculate the electronic energy of molecules to the precision of the thermal energy at room temperature. The results are used to predict reaction rates, equilibrium geometries, transition states, and many other properties.

\section{Group theory and representation theory }
In this section, I will try to give a synopsis of elementary group theory and, while formal for physics, the mathematics will be no deeper than an introductory course on representation theory. Whether one is already familiar with these topics or a newcomer, it will be useful to peruse Table~\ref{tbl:group-notation} for the notation we will be using. 

\begin{table}[th]
    \centering
    \begin{tabular}{cl}
         \hline
         \hline
         $G$          & a finite set of objects and a binary operation, $*: G\times G \mapsto G$, that together\\
                      & satisfy the closure, identity, inverse, and associativity conditions.\\
                      
        $\mu$         & the group action dictates the action of the group on a vector space; $\mu:G\times \vX\mapsto \vX$\\
         $e$          & is the identity element of a group.\\
         $h$          & the finite number of elements in group $G$.\\
         $g$, $g'$    & are generic elements of group $G$.\\
         $gg'$        & is the group element to which $g * g'$ corresponds.\\
         $D^g$        & is a linear representation of $g\in G$.\\
         $\vX$        & is the vector space with basis vector $\{\ej k\}$ that carries the representation $D^g$ of $g\in G$.\\
         $d(\lambda)$          & is the dimension of carrier space $\vX$ for irrep. $\lambda$.\\
                      & Thus, the matrix $D^g$ is $[d\times d]$ when $\vX$ carries the representation.\\
         $\lambda$    & is the label for an irreducible subspace. \\
         $[\lambda]$    & is the Young frame for irreducible subspace $\lambda$ of either the permutation or unitary group.\\
        $T_j^{[\lambda]}$ & The tableau corresponding to the $j$th basis state of irrep.  $\lambda$.\\
        $E^{[\lambda]}_{jj}$ & is a Young project for the $j$th basis state of irrep. $\lambda$.\\
         $D(\lambda)$ & is the set of matrices $\{ D^g(\lambda)\}_{g\in G}$. \\
         $D^g(\lambda)$& is the $\lambda$ representation of $g\in G$ of dimension $[d(\lambda) \times d(\lambda)].$\\
         $\PP^{(\sigma)}$ & the natural representation of permutation $\sigma$; dimension $[h \times h]$\\
         $\mathcal{S}$ & symmetrizer\\
$\mathcal{A}$ & antisymmetrizer\\
$S_n$ & the symmetric group of the permutations of $n$ objects\\
$\sigma,\tau\in S_n$ & are generic permutations from $S_n$; specific permutations are denoted with cycle notation.\\
         \hline
         \hline
    \end{tabular}
    \caption{Table of notation and conventions for group theory and for representation theory.}
    \label{tbl:group-notation}
\end{table}
This discussion of group theory and representation theory is standard introductory material; hence the lack of references. That said, we will provide the reader with guidance on where to go for further reading in Section \ref{sec:further-reading-grps}.
We begin with groups and the group action in Section \ref{sec:grps}. Then we turn toward representations in Section \ref{sec:repr} and close with Young tableaux in Section \ref{sec:young}.

\subsection{Groups and group actions}\label{sec:grps}

The definition of a group $G$ is a set of objects along with a binary composition operation $* : G \times G \mapsto  G$ that obeys the following rules: closure, identity, inverse, and associativity. My personal mnemonic is ``CIIA watches groups.''
\begin{itemize}
    \item \textbf{Closure}: The first rule states that the binary composition of two group elements results in another object within the group.
    \item \textbf{Identity}: The identity rule requires that all groups have an element $e\in G$, which acts as an identity under the composition operation such that $g * e = e * g = g$. 
    \item \textbf{Inverse}: The inverse rule states that if $g$ is an element of the group, then $g^{-1}$ is also an element. That is: if $g\in G$ then there exists $g^{-1}\in G$ such that $g*g^{-1}=e$.
    \item \textbf{Associativity}: Finally, the last rule requires the composition operation to be associative, i.e. $g * (g' * g'') = (g * g') * g''$.
\end{itemize}   

The definition of a group that we have given is sufficiently abstract that many mathematical and physical constructs are captured by this definition. 
You may also think of a finite group with $h$ group elements as being defined by its \emph{multiplication table}.  The table defined by the group's binary operation, $*$, is an $[h\times h]$ matrix defined by $\bra{g} T \ket{g'} = g*g'\in G$. \exercise{Exercise: Give the $[6\times 6]$ multiplication table of the $S_3$ group of permutations.}\exercise{Exercise: What are the constraints on a group multiplication table that allow it to be consistent with each of the CIIA properties of a group?}

When discussing a specific group, it is often sufficient to give the \emph{generators of the group}. The generators of a group, $\langle a,b,\dots, c \rangle$, are a subset of $a,b,..,c\in G$ where all elements of $G$ can be reached by some arbitrary length products of the generator elements. This is akin to the completion of the Lie algebra by repeatedly using the commutators of the control Hamiltonians. Additionally, the \emph{presentation of the group} is the list of generators and some defining relationships sufficient to reproduce the multiplication table.

\begin{figure}
    \begin{center}
        \begin{tabular}{cc}
    \includegraphics[width=.2\linewidth]{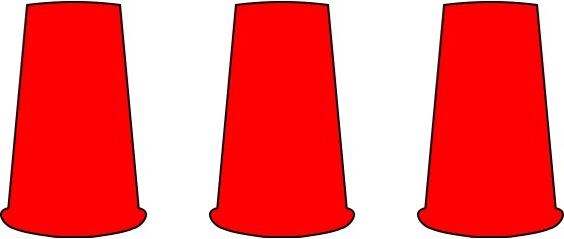} 
    &
\includegraphics[height=13ex]{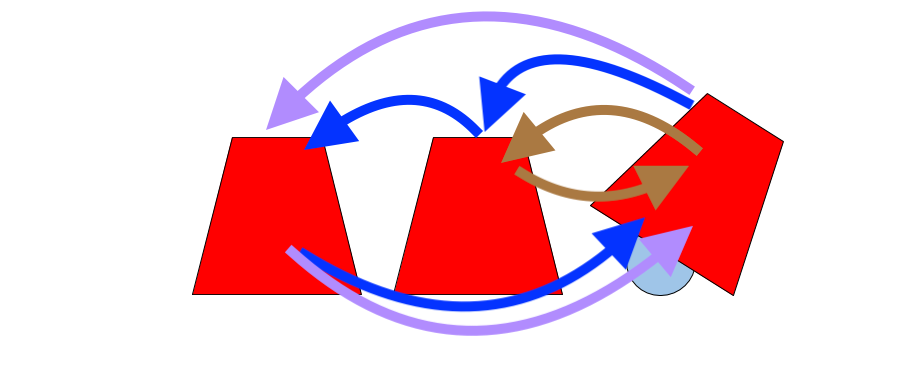}
    \\[1ex]
        Three cup Monte& 
        Active cup permutations
        \end{tabular}
        \end{center}
        \caption{
        The Monte game is played where an adversary (the hustler) exchanges the cups around in front of the player (the mark). After a rapid succession of swaps, the resulting permutation $\sigma$ has been applied: $\PP^{\sigma} \cdot \ket{k} =  \ket{\sigma(k)}$. The player is then asked to guess where the ball is after the permutation. That is, they must select $\ket{\mathbf{\hat e}_j}=\ket{\sigma(ball)}$ to win the game.
        }
        \label{fig:Monte}
\end{figure}

The elements of the group act on an object or on a collection of objects. Consider the Monte game in \figref{fig:Monte}. There, a series of active permutations are performed on the cups by the adversary. The game player must attempt to follow the permutation applied and select the appropriate cup after the adversary has completed their mixing procedure.

More formally, the group action gives us a way of distinguishing the group from the objects that the group acts on.  Suppose that we have a set of points or vectors, $\vX$, that is acted on by the elements of the group. Then we define the \emph{group action} as $\mu: G\times \vX\mapsto \vX$. Note well that the group action is distinct from the binary operation of the group. The group product is $G \times G\mapsto G$ where as the group action is $G\times \vX \mapsto \vX$.
The group action must satisfy
\begin{align}
    \mu(e,A) &= A\\
    \mu(g*g',A)& = \mu ( g , \mu (g',A))
\end{align}

\subsection{Representation theory}\label{sec:repr}

We begin by introducing matrix representations of group elements and some examples. We will will further divide representations into reducible and irreducible and give examples of both. The full list of representations we will introduce throughout this section is given in Table~\ref{tbl:reprs-notation}. Next, we start with the group action on vector spaces.

\begin{table}
\centering
\begin{tabular}{ccc}
\hline
\hline
 & Name & Dimension \\
 \hline
alt & Alternating & 1\\
triv & Trivial & 1\\
reg & Regular & $h=n!$\\
nat & Natural & $n$\\
std & Standard & $n-1$\\
\hline
\hline
\end{tabular}
\caption{Table of representations of $S_n$ group of permutations.}
\label{tbl:reprs-notation}
\end{table}

Now that we have the group action, we can start creating \emph{matrix representations} of the group elements. \emph{Representation theory} is the classification and characterization of vector subspaces that groups can act upon and the appropriate matrices to represent $g\in G$ within these subspaces. It should be noted at the outset that we are developing theory starting from the definition of a group and arriving at representations of the group; however, in practice, often one already has a linear representation of some group's action and wants to know what group the representation came from. Similarly, when we look at an object as a set of vectors/points describing it, then we can get the representation of the symmetries of the object and work backward to understand the symmetry group.

Let us start the discussion with a group $G$ and the group action on set $\vX=\{ \ej j\}^d$. Then the group action gives us a way to find matrix representations of the group operations. We will denote the group action on these vectors as $\mu(g,\ej j) = \ket{ g\cdot \mathbf{\hat e}_j}$. If action of the group results in a vector that is still in the subspace spanned by $\vX$ then we have $\mu(g,\ej j) = \sum_k^d c_k \ej k$.  In this $d$-dimensional subspace, we can then form the $[d\times d]$ matrix $D^g$ as 
\begin{equation}
    D^g = \sum_{j}^d\sum_k^d \edj k  \mu\left(g, \ej j \right)  \; \ej k  \edj j
    \label{eq:grp-action-repr}
\end{equation}
The group action allows us to obtain the matrix representation of $g$ as
\begin{eqnarray}
 \sum_i^d D^g_{kj} \ej k = \ket{ g\cdot  \mathbf{\hat e}_j}\\
  D^g_{ij}  =\langle\mathbf{ \hat e}_i \ket{ g\cdot  \mathbf{\hat e}_j}
\end{eqnarray}
The set of vectors, $\vX=\{ \ej j\}^d$, which we use to write down a specific representation, $D^g$, is said to \emph{carry the representation}.

The matrix $D^g$ represents the group element $g$ as a linear operator. The set of $[d\times d]$ matrices $\{D^g : g\in G\}$ is a \emph{matrix representation} of the group as long as $D^g D^{g'} = D^{g*g'}$. That is, the products of the matrices follow the composition rule of the group elements. More formally, this is called a \emph{group homomorphism}. A function is called a homomorphism if the group product is preserved after the function is applied, that is, $f(g)\cdot  f(g')=f(g*g')$ for all $g,g'\in G$.

Only the trace of the representations is considered in \emph{character theory}. In the context of group theory, the \emph{character of a representation} is given by the trace of the representing matrix: $\chi_g=\Tr[D^g]$.

Uninitiated readers should be wary of overgeneralizing the definitions given so far. First, note that the dimension of the carrier space $d$ is not fixed \textit{a priori} and can range from $1$ up to $h$ (the number of group elements) or beyond. In the following, we will consider $n$-dimensional vector spaces and $d$-dimensional subspaces, and hence $d\leq n$. Regardless of the value of $d$, again, the only requirement is that the $[d\times d]$ representations follow the multiplication table of the group. 

\paragraph{Trivial representation}
The first example of a representation is the \emph{trivial representation}. It is a $d=1$ representation where $D^g=1$ for all $g$. This always satisfies the multiplication table for any group since
\begin{equation}
\begin{array}{rccccc}
    &D^g(\text{triv})&\cdot & D^{g'}(\text{triv})& =&D^{g*g'}(\text{triv})\\
   =&1 &\cdot& 1 &=& 1
\end{array}
\end{equation}
The trivial representation exists for all groups.  

A representation is called \emph{faithful} if each group element maps to a different representation. More formally, the representation mapping, $D$, is injective i.e. $D^{g} \neq D^{g'}$ whenever $g\neq g'$.  Therefore, the trivial representation is, in a sense, the least faithful since all group 
elements are mapped to the same representation. 

The subspace that carries this trivial representation is straight-forward to construct and will connect to other areas readers may have encountered. Specifically, \emph{symmetrization} can be used to project any vector from any given carrier space into the one-dimensional subspace that carries the trivial representation. For any group $G$, we can symmetrize a state by acting with the sum of all group elements. For any representation, $D^g$, we  project states into the trivial subspace using:
\begin{eqnarray}
\mathcal{S}&=&\sum_{g\in{G}} D^g
\label{eq:symmetrizer}
\end{eqnarray}
The action of $\mathcal{S}$ on a state will project the state into the subspace that carries a trivial representation of the group.\exercise{Exercise: Show that the symmetrizer is proportional to a projector i.e. $\mathcal{S}^2=c \mathcal{S}$ with $c\in \reals$. }

\paragraph{Regular representation}
Next, we turn to the \emph{regular representation}. This representation can also be defined for all groups. We reach this representation by considering the group action acting on the group itself: 
$$\mu(g,g') = g * g' = gg'$$ 
This representation has $d=h$ and is given by a permutation matrix that permutes the group elements according to the multiplication table such that:
\begin{equation}
    D^g(\text{reg}) \ket{g'} = \ket{gg'}
\end{equation}

\paragraph{Reducibility of representations}
Now 
we are in a position to consider an important area of representation theory: \emph{irreducible representations}. 
We can start by considering the carrier space of a representation. Consider a representation of a group labeled by $\lambda$:  $\{D^g(\lambda)\}$. Suppose that this representation is carried by vector subspace $\vX=\text{span}\{\ej j\}^d$ of an $n$-dimensional vector space.  Even though representation $D^g(\lambda)$ is $[d \times d]$, we can embed it into the full $n$-dimensional vector space. This can be done using a direct sum of $D^g$ with an identity matrix of size $[(n-d)\times (n-d)]$. Then the $[n \times n]$ matrices still form a valid representation of $G$.

The direct sum of matrices $A$ and $B$ is
\begin{equation}
    A\oplus B = \left[\begin{array}{c|c} A & \mathbf{0} \\
    \hline
    \mathbf{0} & B
    \end{array}
    \right]
\end{equation}
If the dimensions of $A$ are $[n\times n']$ and $B$ has dimension $[m\times m']$ then $A\oplus B$ has dimension $[(n+m)\times (n'+m')]$.\exercise{Exercise: Prove that the direct sum of two unitary matrices is unitary.} 

We can take this idea in the reverse direction and start with the $n$-dimensional vector space carrying a $[n \times n]$ representation. Within this $n$-dimensional vector space there is a set $d$ of vectors spanning subspace $\vX$ that allows us to decompose $D^g$ into the representation on $\vX$ and a representation on its complement $\vX^c$.\exercise{Exercise: If $\vX$ carries a representation, it is reasonable to guess that the complement of the space, $\vX^c$, also carries a representation. Prove that this is the case or provide a counter-example.}

Suppose subspace $Y\subset \vX$ carries a \emph{reduced representation}, then the group action on items within $Y$ only results in other items still within subspace $Y$ i.e. the subspace $Y$ is \emph{invariant under the action of $G$}. When vector space $\vX$ contains such an invariant subspace, then $\vX$ is said to be an \emph{reducible subspace}.  If there is no smaller subspace of $Y$ that is also invariant under the action of the group $G$, then $Y$ is said to be an \emph{irreducible subspace}. The representation that an irreducible subspace carries is called an \emph{irreducible representation}. 

The trivial representation is an irreducible representation (there is no smaller subspace since it is already one-dimensional). The regular representation contains all irreducible representations. Moreover, the number of copies of each irreducible representation, $\lambda$, is the dimension, $d(\lambda)$, of that irreducible subspace. In symbols, 
\begin{equation}
D^g(\text{reg}) =\bigoplus_\lambda^{irreps} \left( \oplus_{i=1}^{d(\lambda)}D^g(\lambda) \right)
\label{eq:regular-dcomp}
\end{equation}
By considering the dimensions of these representations, we also have the consequence
\begin{align}
    h &= \sum_{\lambda} d(\lambda)^2 
\end{align}
We can also consider character theory and apply these concepts directly to the trace of the representations:
$\chi_g(\text{reg})=\sum^{irreps}_\lambda \chi_g(\lambda)d(\lambda)$. In the next section, we use group $S_3$ to illustrate examples of representations.

\paragraph{Examples using the permutation group $S_3$; natural and standard representations}
Given the definition of the regular representation, every finite group is isomorphic to a subgroup of the permutation group. Therefore, understanding the structure of permutations helps pave way not only for the study of identical particles but also for illustrating many concepts in group theory more broadly. The representations of permutations give a nice chance to illustrate the idea of reducible and irreducible representations. 

Let us consider the group $S_3$ explicitly. We have generators $\langle (12), (23)\rangle$ and $h=6$ elements total. They are 
$$\{e, (12), \;(23), \;(12)*(23), \;(23)*(12), \; (12) * (23) * (12)\}$$
And we will note that $ (13) = (23)*(12)*(23)= (12) * (23) * (12) $. 
We define a matrix representation of permutations using the following, seemingly natural definition:
\begin{align}
    \PP^{\sigma}=\ej{\sigma(k)}\edj k
\end{align}
This representation is called the \emph{natural representation} where the consider the permutations of $n$ items acting on the basis vectors of an $n$-dimensional vector space such that:
\begin{equation}
\mu((12),\begin{bmatrix}v_a\\ v_b \\ v_c \end{bmatrix})=\begin{bmatrix}v_b \\ v_a \\ v_c\end{bmatrix}.
\end{equation}
Then the representation of $S_n$ is a set of $[n\times n]$ matrices. The natural representation for $S_3$ is 
\begin{equation}
    \left\{\PP^{(12)}=D^{(12)}(\text{nat})  
    =\begin{bmatrix} 0 & 1 &0\\1&0&0\\0&0&1\end{bmatrix},
    \PP^{(23)}=
    D^{(23)}(\text{nat})
    =\begin{bmatrix} 1 & 0 &0\\0&0&1\\0&1&0\end{bmatrix}\right\}\label{eq:natural-repr}
\end{equation} 
The natural representation is a reducible representation because it also contains the symmetric subspace. The \emph{standard representation} is the irreducible representation formed from the remainder of the natural representation after projecting out the symmetric subspace. Establishing the representations on the remaining two-dimensional space is left as an exercise.\exercise{Exercise: Compute the standard representation of $S_3$. Start by orthogonalizing the two-dimensional space orthogonal to the vector $\vec u$ in \eqref{eq:unif}. (\textit{Hint:} Both negative and fractional numbers will appear in the representation.)  Show that the matrices obtained remain a valid representation of $S_3$ e.g. check that $D^{(12)}(\text{std}) ^2 = D^{(13)}(\text{std})^2 = \id$.} 

\paragraph{One-dimensional representations}
Next, we consider the symmetrized state. As introduced earlier, the symmetrized state will carry the \emph{trivial representation}. For $S_3$, the carrier of the trivial space is
\begin{equation}
\ket{u} = \begin{bmatrix}1\\1\\1\end{bmatrix}
\label{eq:unif}
\end{equation}
Note that both generators of the group given in \eqref{eq:natural-repr} leave $\ket u$ invariant. It then follows that any product of the generators will also leave $\ket u$ invariant.

Next, we turn to the \emph{alternating representation}. It is another one-dimensional representation for permutations. The parity of the transpositions, i.e. even or odd, forms a valid representation of the permutation group. This representation is called the \emph{alternating representation}. 
\exercise{Exercise: The projector into antisymmetric spaces can be written as $\mathcal A = \sum \sgn(\sigma) \;\ket{\sigma(x)}\bra{x}$. }

We will need to define the sign of a permutation, $\sgn(\sigma)$. If the permutation $\sigma$ is composed of an even number of transpositions, as in $\sigma=(12)*(13)$, then $\sgn(\sigma)=+1$. If instead $\sigma$ is composed of an odd number of transpositions e.g. $\sigma=(12)*(23)*(12)=(13)$, then $\sgn(\sigma)=-1$. The generators of $S_3$ in the alternating representation are simply: $D^{(12)}(\text{alt})=-1$ and $D^{(13)}(\text{alt})=-1$.

\paragraph{Regular representation of $S_3$}
The regular representation of $S_3$ can be tedious to construct but we shall do the example here. We first need to select an ordering of our vector space. Let's use
\begin{equation}
    \ket{x_0} = 
    \begin{pmatrix}
        abc\\
        bac\\
        cab\\
        cba\\
        bca\\
        acb
    \end{pmatrix} = 
    \begin{pmatrix*}[r]
        e \cdot abc\\
        (12)\cdot abc\\
        (13)*(12)\cdot abc\\
        (13) \cdot abc\\
        (12) * (13) \cdot abc\\
        (23) \cdot abc
    \end{pmatrix*}
\end{equation}
as the basis state for our vector space. Note that we have used the common shorthand $\sigma \cdot abc = \mu(\sigma, abc)$. Now, we can consider the action of $\sigma=(12)$ on this state.
\begin{equation}
    \mu(\sigma,\ket{x_0})  = 
    \begin{pmatrix*}[r]
        (12)*e \cdot abc\\
        (12)*(12)\cdot abc\\
        (12)*(13)*(12)\cdot abc\\
        (12)*(13) \cdot abc\\
        (12)*(12)*(13) \cdot abc\\
        (12)*(23) \cdot abc
    \end{pmatrix*}
    =
    \begin{pmatrix}
        bac\\
        abc\\
        acb\\
        bca\\
        cba\\
        cab
    \end{pmatrix} 
    =\ket{x}
\end{equation}
Next, we write down the representation matrix that transforms as $D^{(12)}(\text{reg})\ket{x_0}=\ket{x}$. By inspection:
\begin{equation}
    D^{(12)}(\text{reg})= \begin{pmatrix}
        0 & 1 & 0 & 0 & 0 & 0\\
        1 & 0 & 0 & 0 & 0 & 0\\
        0 & 0 & 0 & 0 & 0 & 1\\
        0 & 0 & 0 & 0 & 1 & 0\\
        0 & 0 & 0 & 1 & 0 & 0\\
        0 & 0 & 1 & 0 & 0 & 0
    \end{pmatrix}
\end{equation}
Constructing another generator is left as an exercise. \exercise{Exercise: Give the regular representation of $\sigma=(13)$ for $S_3$.} Since this is the regular representation, we should have the decomposition $D^g(\text{reg})= D^g(\text{triv}) \,\oplus \,D^g(\text{alt})\,\oplus\, D^g(\text{std}) \,\oplus\, D^g(\text{std})$. The standard representation is two-dimensional so it appears twice in the regular representation per \eqref{eq:regular-dcomp}.\exercise{Exercise: Compute the irreducible representations of the regular $[6\times 6]$ representation for elements $(12)$ and $(23)$ of $S_3$. It is suggested to begin this exercise by removing the $[111111]^T$ subspace, then removing the appropriate one-dimensional vector that carries the alternating representation. Finally, one must muscle the remaining $[4\times 4]$ block into two copies of the two-dimensional irreducible subspace of $S_3$.}

\subsection{Conjugacy classes, Young frames, and Young tableaux}\label{sec:young}

In this section, we continue the discussion of representation theory with facts concerning the Young diagrams and the classification and representation of permutations. Three years after JJ Thomson discovered the electron, Alfred Young introduced his namesake diagrams to think about representations of the permutation group of $n$ objects.

The Young frames are a way to label the components of the symmetric group. We will use the term \emph{Young frames} to refer to the empty diagram, \emph{Young tableau} to refer to the Young frame filled with numbers, and \emph{Young tableaux} is the plural of Young tableau.
\emph{Young frames} are equivalent to partitions of $n$ objects, $[\lambda] = [r_1, r_2, \dots , r_k]$, with $\sum r_i=n$ and $r_i\geq r_{i+1}\geq0$.  Each $r_j$ indicates the number of boxes in the $j$th row of Young frame $[\lambda]$. 

The Young frames label the conjugacy classes of permutations. In common group theory parlance, two group elements $a,b\in G$ are in the same \emph{conjugacy classes} if $a=g * b * g^{-1}$ for some $g\in G$.  

A \emph{cycle} in a permutation is the elements of the orbits of an index under repeated action of the permutation. An example of a three cycle is $\sigma$ such that $\sigma(1)=2$, $\sigma(2)=3$ and $\sigma(3)=1$. We write this permutation in cycle notation as $\sigma=(123)$. A second example is $\tau=(134)(25)$ with indices mapped as
\begin{equation}
\begin{array}{c|ccccc}
    i & 1 & 2 & 3 & 4 & 5\\
    \tau(i) & 3 & 5 & 4 & 1 & 2
\end{array}
\end{equation}
The cycle structure for the second example is a three-cycle and a two-cycle. 

\begin{figure}
        \centering
  \includegraphics[width=.5\linewidth]{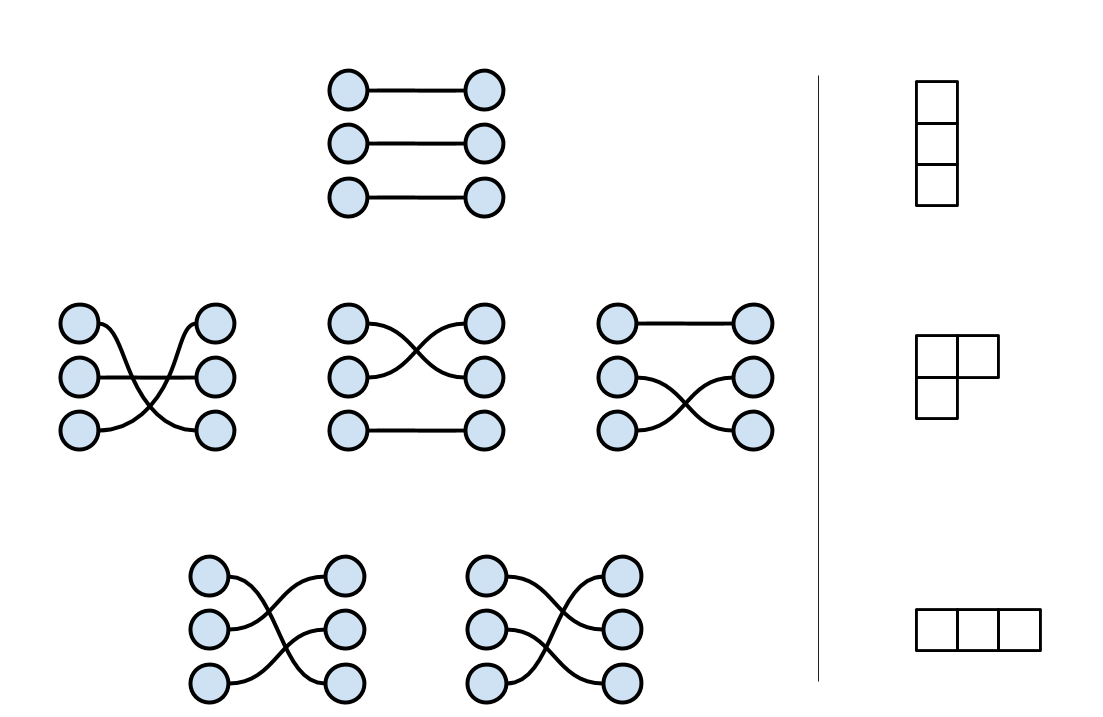}
        \caption{Graphical representation of the $S_3$ group of permutations. The permutations are grouped by their cycle structure. The cycle structures are labeled by Young tableaux on the right.
        The matrix representations of these graphics depends on if one reads left-to-right or right-to-left. The same issue arises in other graphic representations of composition, such as the quantum circuit model.}
        \label{fig:s3}
    \end{figure}

The cycle structure of a permutation is preserved under conjugation by other group elements. Thus, we can label the conjugacy classes by the cycle structure. 
For permutation $\sigma$, conjugation by any other group element will preserve the cycle structure of $\sigma$. Thus, the Young frames also symbolize the cycle structure with the length of each row indicating the cycle lengths within the permutation. Examples of Young frames for $S_3$ are shown in \figref{fig:s3}.

The Young frames also label the irreducible subspaces
of the \emph{unitary group}. Here, elements of the unitary group $U\in U(n)$ are given by $n$-fold tensor products of a local unitary $u$: 
\begin{equation}
\{U=u^{\otimes n}: u^\dag u =\id \}
\end{equation}
In general, this captures the idea of a local 
operation applied uniformly to $n$ disjoint systems.

The \emph{Schur-Weyl duality} states that the unitary and permutation groups maximally commute. Consider permutations $D^{(12)}$ acting on operators $A\otimes B$ as 
\begin{equation}
    D^{(12)}(A\otimes B) 
\end{equation}
Since $u$ acts identically in each subspace such permutations will have no effect on $U=u^{\otimes n}$. For example, with $n=2$,
\begin{equation}
    D^{(12)}(u\otimes u) 
\end{equation}
This is the essence of the duality between the two groups. 
As a result of this duality, the irreducible subspaces of the permutation group can also be labeled by Young frames.

The basis functions of each irreducible representation are obtained by the standard Young tableaux. Here, the rows and columns of a Young tableau are filled with numbers $1$ through $n$ such that the numbers increase across the rows and increase down the columns.  For example, with $\lambda_1=[21]$, $\lambda_2=[3]$ and $\lambda_3=[111]$,  the standard tableaux are
\begin{align}
	T^{[\lambda_1]}_1
 &=\young(12,3) &
 T^{[\lambda_1]}_2
 &=\young(13,2) &
 T^{[\lambda_2]}
 &= \young(123) &  
 T^{[\lambda_3]}
 &=\young(1,2,3) 
	\label{eq:examples-s3}
\end{align}

The projection into these subspaces is done by translating a tableau with $n_r$ rows and $n_c$ columns into the \emph{Young operator} $E^{[\lambda]}_j$ that antisymmetrizes the columns and symmetrizes the rows.  Here $N=\mathcal{A}_1\mathcal{A}_2\dots\mathcal{A}_{n_c}$ with $\mathcal{A}_i$ corresponding to the antisymmetrizer for the elements of the $i$th column and $P=\mathcal{S}_1\mathcal{S}_2\dots\mathcal{S}_{n_r}$ with $\mathcal{S}_j$, the symmetrizer for the $j$th row. 

Each standard Young tableaux labels a basis vector of the irreducible subspace. Given a standard tableau $T_A^{[\lambda]}$, we can write down the projection operators, $E^{[\lambda]}_{AA}=\mathcal{S}_{rows(T_A)}\mathcal{A}_{cols(T_A)}$ and transfer operators $E_{AB}^{[\lambda]}=E_{AA}^{[\lambda]}P_{T_A\leftarrow T_B}$.

For some examples:
\begin{align}
    E^{[\lambda_1]} &= (e-(13)) (e+(12)) & E^{[\lambda_2]} &= \mathcal{S} & E^{[\lambda_3]}&=\mathcal{A}
\end{align}
\exercise{Exercise: For $S_3$ show that the $D(std)$ standard representation corresponds to the group action on the $\{T_j^{[\lambda_1]}\}$ carrier space.} 
We now turn toward
connecting various mathematical phrasing of fermions in a variety of contexts ranging from chemistry to condensed matter and statistical mechanics. This will also allow us to appreciate much of the literature on the simulation of bosons and fermions in computational settings from chemistry and physics to quantum information.

\section{Identical  particles}
Now, we turn toward quantum identical particles. There are two one-dimensional representations of the permutation group and both allow us to create invariant quantum states that will not change outcome probabilities under the action of particle  permutations. 

\subsection{First quantization}
Before moving directly to fermions or bosons, let's first consider the generic multi-particle situation. Let $d$ be the number of locations available to a single particle and $N$ denote the number of particles. It then follows that there are $d^N$ configurations of the $N$ particles.  Letting $x_j=1\dots d$ then $\Psi_{x_1x_2\dots x_N}\in \mathbb{C}$ is just a tensor with $N$ indices each taking values from 1 to $d$. This presentation of the quantum pure state $\Psi$ is often called first-quantization in the quantum simulation literature.

Let's begin with the $N=2$ situation. We can write the wave function as a matrix $A_{ij}=\Psi_{{x_1=i},{x_2=j}}$. If this matrix is symmetric $A=A^T$ or antisymmetric $A=-A^T$ then we get bosons and fermions respectively:
\begin{align}
    A_{ij}=A_{ji} &\iff \Psi_{x_1 x_2}= \Psi_{x_2 x_1} && \text{Bosons}\\
    A_{ij}=-A_{ji} &\iff \Psi_{x_1 x_2}= -\Psi_{x_2 x_1} && \text{Fermions}
\end{align}
In the two cases, note that the transpose of $A$ is serving as the particle swap operator.

Of course, there are matrices which are not symmetric nor antisymmetric. However, for all matrices $A = A_{sym} + A_{asym}$ where $A_{sym}=(A+A^T)/2$ and $A_{asym}=(A-A^T)/2$ are the projections of $A$ into the two matrix subspaces.

\subsection{Second quantization}
Second quantization takes into account the fact that symmetric and antisymmetric matrices/tensors have a large amount of redundancy. The matrix condition $A=\pm A^T$ means the lower left triangular portion is irrelevant. The same idea extends to multi-index tenors that are completely antisymmetric or completely symmetric. 

In second quantization, we have basis states labeled by the configuration of orbital occupancy. We explicitly impose an ordering on the local states with each location specifically identified and labeled. The particles, by contrast, are no longer explicitly tracked. Instead, we place an unlabeled particle into a labeled location. The locations are called orbitals.  

\emph{Orbitals} are, under the broadest definition, a function mapping from the coordinate space $\vX$ of a single particle to complex numbers i.e. $\phi: \vX \mapsto\cplxs$. Consistent with our earlier discretization, $\dim(\vX)=d$.
\footnote{If one is considering real space, for analytic purposes, consider a sequence of finite dimensional space and take the limit as $M\rightarrow \infty$. For numerical purposes, one may consider $M=2^{64}$ as a reasonable  representation of the position coordinate. In practice, $d$ is usually less than 1000 for fermion simulation of molecules.  }
There are several terms in use that refer to the same concept: \emph{sites}, \emph{modes}, \emph{locations}, etc.  Any other name that refers to such a single-particle functions is equivalent. One may hear all of these terms used interchangeably to refer to  one-particle functions.

\subsection{Operator algebras}
The resulting states will be completely antisymmetric or completely symmetric so long as we traverse the $d^N$ dimensional state space of $N$ particles correctly and consistently. We will use collections of \emph{particle operators} $$\left\{\{ a_j\in [2^d\times 2^d]| j=1\dots d\} \cup  a_j^\dagger\in [2^d\times 2^d] | j=1\dots d\}\right\} $$ to navigate a (usually) smaller space of dimension $2^d$ rather than $d^N$. 

In quantum theory, we often use the term \emph{operator} to indicate a linear operation on states even in finite dimensions. In finite dimensions, the operator will have a representation as a matrix and the terminology runs parallel to the earlier discussion of permutations as  operations and matrix $D^\sigma$ as its representation.

Similar techniques arise in the ladder methods for differential equations, orthogonal polynomials, and in the theory of Lie algebra. In all contexts, including the present, the ladder operators transform between eigenstates of a target operator.  Here, the raising and lowering operators change the eigenvalue of the local number operator $\hat n_j=a_j^\dagger a_j$. 

The vacuum state is defined as the state that is in the kernel of all $\{a_j\}$ i.e. 
\begin{equation} 
a_j\vac =0\label{eq:vac}
\end{equation} 
regardless of the specific algebra obeyed by operators $a_j$. The vacuum state, $\vac$, is the normalized state  of dimension $[2^d\times 1]$ that represents the system with no particles. The \emph{creation} or raising operators and the corresponding \emph{annihilation} or lowering operators act on state $\vac$ to insert or remove identical particles into corresponding modes.

For fermions, the raising and lowering ladder operator  must follow the \emph{canonical anticommutation relations} for fermionic operators \begin{align}
&[a_p, \;a_q]_+= a_pa_q+a_qa_p=0,& &[a_p, \;\ad_q]_+=a_p\ad_q+\ad_qa_p=\delta_{pq} \id.
\label{eq:CAR}
\end{align}
This canonical anticommutation relation defines Dirac fermions. \exercise{Show that the Pauli exclusion principle: ``that no two fermions can occupy the same state" follows directly from \eqref{eq:CAR}.}
Similarly, for bosons the operators follow a commutator algebra whereby
\begin{align}
[a_p,a_q] &= a_p a_q - a_qa_p =0&
[a_p, a_q^\dagger ]&=a_p a_q^\dagger - a_q^\dagger a_p =\delta_{pq} \id
\label{eq:CCR}
\end{align}

Using the relations in \eqref{eq:CAR} and \eqref{eq:CCR} requires the order of the operators to be fixed. This is imposed by the linear indexing of the orbitals regardless of the spatial of the orbitals layout. This leads to complications when discussing the locality of the interaction since orbitals that are spatially close may not be numbered sequentially. 

Acting on the vacuum state with a string of $N$ distinct creation operators yields $N$-body event states for identical particles.  Each creation operator strings is labeled by occupation number string 
$$
\vec K=(K_1, K_2, \dots ,K_d)
$$ 
with $K_j\in \{0,1\}$ for fermions and $K_j\in\{0,1,\dots N\}$ for bosons. The corresponding operator string is
\begin{align}
\hat K=\prod_{j=1}^d (a_j^\dagger)^{K_j}
\end{align}
We also say that $\ket{K}$ is the event state corresponding to finding occupancy pattern $\vec K$ upon measurement.  

The space of states generated this way is called \emph{Fock space}. Then \emph{Fock states} are
written as
\begin{align}
    \ket{K}&= \hat K\vac =  \prod_{j=1}^d  \left(a_{j}^\dag\right)^{K_j}\vac\\&= a_{k_1}^\dagger a_{k_2}^\dagger a^\dagger_{k_3} \dots a^\dagger_{k_N} \vac
\label{eq:wf-second}
\end{align}
Here $k_j$ is the index of the $j$th non-zero entry of $K$.

We can break the Fock space into fixed particle number sectors, $\mathcal{F}_N=\{\ket{K} : \sum K_j =N\}$, where for a given value of $N$, there are $\binom{d}{N}$ states corresponding to possible $N$-particle event states that can be realized upon measurement. 

For fermions, $K_j\in\{0,1\}$ and, $1\leq k_i<k_j\leq d$ for all $i < j \leq N =\sum_j^d K_j=|K|$. In a fixed-particle number sector with $N$ particles there are $\binom{d}{N}$ states.\exercise{The binomial coefficient $\binom{m}{n}$ is given by 
$$\frac{m!}{(m-n)!n!}$$
\begin{itemize}
    \item Show for $m\gg n$, the binomial coefficient  $\binom{m}{n}$ can be approximated as $m^n/n!$ 
    \item Show for $m\gg n$, binomial coefficient $\binom{m+n-1}{m-1}$ is also approximately $m^n/n!$
\end{itemize}
These limits serve to confirm the initial understanding that we are only storing a ``corner'' of the full tensor since the rest of the tensor is redundantly related through the symmetries the state satisfies.
}
For bosons, $K_j\in\{0,1,2,\dots N\}$ and $1\leq k_i\leq k_j\leq d$. In the $N$ boson fixed-particle number sector, there are $\binom{N+d-1}{d-1}$ states. \exercise{Use the ``stars-and-bars" method to count the number of states in a fixed-particle bosonic sector. Start with $N$ stars in a list, then partition the list into $d$ bins by inserting the correct number of bars. Count the number of ways the bars can be inserted to arrive at correct binomial coefficient.}

In the fermonic setting, the total Fock space can be written $\mathcal{F}=\bigoplus_{N=0}^d \mathcal{F}_N$.  As a reminder:
$\sum_{N=0}^d \binom{d}{N} =2^d$. This fact serves as a confirmation that the dimensions assigned to each subspace of fixed particle number are indeed. This shows that the dimensions are consistent since 
there are $2^d$ binary strings of length $d$. 

Going beyond Dirac fermions, we consider any set of $2d$ operators, $\alpha_1,\dots \alpha_{2d}$
to be fermionic if they satisfy the generalized anticommutation relation:
\begin{align}
\alpha_i\alpha_j +\alpha_j\alpha_i = S_{ij} \id
    \label{CAR-general}
\end{align}
As long as the $[2d \times 2d]$ matrix $S$ satisfies $S=S^T$, then these operators represent a fermionic algebra.  If $S$ is diagonal, the fermions are often called \emph{Majorana fermions}.

\section{Outlook and summary}

In this article, we have looked at the mathematics underlying permutations and their representations, consequences for identical particles, and a look at first and second quantization schemes for representing systems of identical particles. The goal is that readers are able to understand their own area of research better as well as springboard into adjacent fields. 

We have been largely agnostic about where these particles appear, thus we have not made an effort to tackle specific applications.  Were we to choose foundational applications in the direction of physics and chemistry, our discussion would continue with the free electron gas and molecular hydrogen. For quantum information science, the theory of non-interacting fermions connects to graph theory and provides a handle on the computational divide between quantum and conventional computation \cite{Terhal2002Mar}.

To discuss operators like the Hamiltonian, minimally, we must introduce notions of energy and of time. A consideration of the time evolution operators of fermionic systems would highlight the distinction between rotating the local orbital basis and rotating the $N$-body basis. Finally, we could continue with a look at the optimization of multi-particle states with respect to some operator. Such optimization procedures form an important area of quantum chemistry, condensed matter, and quantum computing largely dictated by algorithms from density functional theory, density matrix renormalization group algorithms, and variational quantum algorithms in those respective domains.

Lastly, we will mention the braid group. The action of swapping two particles in two-dimensions can be done in two inequivalent ways. In the braid group, the interchange of two identical particles can be done in an over-under or an under-over fashion. Identical particles in two-dimensions carry a representation of the braid group because of this inequivalence of the two exchanges. 

\section{Further reading}\label{sec:further-reading-grps}
Recommended textbook references on identical particles and related mathematics come in several categories. I originally learned group theory as a second year undergraduate at Morehouse College using Ref.~\cite{gallian2021contemporary}.  Ref.~\cite{dummit1991abstract} is a fairly popular mathematics textbook. In the context of physics,  Refs.~\cite{sternberg1995group}, \cite{hammermesh}, and \cite{Tinkham2004Jun} are key references.  In the chemical context, Ref.~\cite{cotton1991chemical} is a standard introduction to group theory, while Ruben Pauncz has written several textbooks, e.g. Ref.~\cite{Pauncz95} is focused on the treatment of spin symmetries in electronic structure. 

The discussion of identical particles is found in most quantum textbooks. In particular, a useful conceptual discussion of postulated exchange statistics is found in Ref.~\cite{ballentine2014quantum} while the later editions of Ref.~\cite{sakurai1995modern} contain an introduction to spin and permutational symmetries. One of the best and most direct introductions to second quantization is found in statistical mechanics textbook Ref.~\cite{pathria2016statistical}.  In quantum chemistry: Ref.~\cite{Szabo96} contains a detailed and extensive introduction to the first quantized representation, while Ref.~\cite{Helgaker00} is phrased almost exclusively in second quantization.  Ref.~\cite{Weedbrook12} is foundational reference for bosons in quantum information. 

\section{Acknowledgments}
I have spent my entire career studying fermions in a wide variety of settings around the world. I would like to highlight particularly useful feedback from C Dowdle. Also, S. Gulania, A. Projanski, W. Wang, and B. Harrison are also gratefully acknowledged for feedback throughout the writing process.

This work was supported by the U.S. Department of Energy Office of Science, Office of Advanced Scientific Computing Research under programs Quantum Computing Application Teams and Accelerated Research for Quantum Computing program, as well as the Department of Energy Grant DE-SC0019374.

\end{document}